\documentclass[12pt]{article}
\usepackage[a4paper, left=2.5cm, right=2.5cm, top=2.5cm, bottom=2.5cm]{geometry} 
\usepackage{amsmath,amssymb,graphicx}
\usepackage{booktabs}

\begin{document}

\title{Metabolic Rate Beyond the $3/4$ Law} 
\author{
	Dorilson Silva Cambui\\
	Governo de Mato Grosso, Secretaria de Estado de Educação\\
	Cuiabá, Mato Grosso, Brazil\\
	\texttt{dorilson.cambui@edu.mt.gov.br}\\
	\texttt{dcambui@fisica.ufmt.br}
}

\maketitle

\begin{abstract}
\noindent
In earlier work, we introduced a discrete Fibonacci-based ontogenetic model in which the metabolic scaling exponent $b(n)$ is treated as a dynamic function of an organism's developmental stage, and we estimated $b(n)$ for selected mammalian species. In the present article, we revisit this framework with a complementary aim. Rather than proposing new parameter estimates or statistical fits, we provide a didactic, step-by-step reconstruction of the derivation that leads from the recursive growth hypothesis to analytical expressions for the stage-dependent exponent $b(n)$. Building directly on these previously obtained exponents, we then incorporate Kleiber's classical result into the model by interpreting the constant $70$ in the law $B \approx 70\,M^{3/4}$ (with $B$ denoting basal metabolic rate and $M$ body mass) as a metabolic "anchoring point". This yields a stage-dependent basal metabolic rate of the form $B(n) = 70\,M^{b(n)}$, which defines an ontogenetic metabolic trajectory linking recursive growth to changes in scaling. We show, at a conceptual level, how this anchored formulation can describe a shift from strongly sublinear behavior at early stages towards an almost linear regime as development proceeds, while still producing basal rates that are compatible, in order of magnitude, with those reported for mammals of different sizes. In this way, the paper offers a self-contained and pedagogical presentation of the model, emphasizing how ontogenetic changes in metabolic rate can be understood through the combined ideas of Fibonacci-like recursion and metabolic anchoring.
\end{abstract}

\section{Introduction}\label{int}

Generally speaking, we can define metabolic rate as the "tempo of life", understood as the speed at which an organism converts resources into energy to sustain all vital functions, from basic cellular processes to the mechanical work of muscles~\cite{delong2010}. This tempo is not the same for all organisms, as it varies with factors such as body size, environmental temperature, and lifestyle. While a giant whale burns energy slowly and steadily, a hummingbird in frenzied flight possesses an extremely accelerated metabolic "engine," requiring constant food intake~\cite{suarez}. This relationship between body size and energy expenditure, classically described by Kleiber~\cite{Kleiber}, allows us to understand how life adapts to different environments and how the balance between energy acquired and energy needed for existence is maintained.

Kleiber's law~\cite{Kleiber} establishes the relationship between basal metabolic rate $B$ and body mass $M$ in the form $B \propto M^{3/4}$, and emerged from a methodological analysis that synthesized experimental data from multiple sources~\cite{savage}. Max Kleiber compiled precise measurements of oxygen consumption (obtained by indirect calorimetry under basal conditions) in mammals spanning a wide range of body masses, from small rodents to cattle and elephants. When he attempted to relate mass and metabolism directly in a standard linear plot, the data produced a curve that was difficult to interpret, with values for small animals clustered at one end and values for large animals spread out at the other. The key to revealing the underlying pattern was to represent the data on a log--log plot, graphing $\log(B)$ as a function of $\log(M)$. In this representation, the previously scattered experimental points aligned approximately along a straight line. This alignment is the mathematical signature of a power law (or scaling law) of the form
\begin{equation}\label{Bclassico}
  B = B_{0} \cdot M^{b},
\end{equation}
where $B$ represents the basal metabolic rate of the animal, $M$ is its total body mass, $B_{0}$ is a proportionality constant that indicates the level of metabolic rate for a unit mass (for example, 1~kg), and $b$ is the scaling exponent, the central parameter of the law. Kleiber applied linear regression analysis to determine the slope of the best-fit line. This slope, denoted by $b$ in the logarithmic equation $\log(B) = b \cdot \log(M) + \log(B_{0})$, corresponds exactly to the scaling exponent in the power law relation (given by Eq. ~\ref{Bclassico}). The value that Kleiber found for this slope was approximately $0.75$, which became known as the law of three-quarters. From the linear fit on a logarithmic scale, Kleiber was able to determine not only the scale exponent $b \approx 0.75$, but also the point where this line crosses the vertical axis (intercept $B_{0}$), which, when converted back from the logarithmic scale to the arithmetic scale, corresponds to a value of approximately $70$.
Thus, the general expression of Kleiber's law, derived directly from his empirical analysis for mammals, can be written in a more complete form as $B \approx 70 \, M^{0.75}$, 
where $B$ is the basal metabolic rate (in kcal$\cdot$day$^{-1}$) and $M$ is the body mass (in kg).
This expression tells us that, on average, each kilogram of a mammal does not "consume" energy strictly in proportion to its mass, but according to the three-quarter power law. It is important to note that the exact value of $B_{0}$ may vary slightly depending on the dataset or taxonomic group, but the $3/4$ exponent was, at the time, regarded by Kleiber as approximately universal among mammals.

While Kleiber's work established that metabolism scales with mass raised to the $3/4$ power, it was the Metabolic Theory of West, Brown and Enquist~\cite{wbe} (WBE), that set out to explain \emph{why} this specific exponent emerges. The WBE model is based on the need to optimize the distribution of resources (nutrients and oxygen) throughout the body. This optimization is achieved through hierarchical transport networks with fractal properties, such as the circulatory and respiratory systems. The theory argues that, to efficiently supply every cell in a three-dimensional body from a single source (such as the heart or lungs), the architecture of these networks must obey certain principles of scale invariance and space filling. The mathematical solution of this optimization problem, taking into account the fractal geometry of the networks, leads to a theoretical scaling exponent of $b \approx 3/4$. In this way, the WBE model offers a mechanistic and theoretical explanation for the law of three-quarters established empirically by Kleiber.

It is important to highlight here that both Kleiber's empirical law and the West, Brown, and Enquist (WBE) theoretical model represent fundamental advances in the description of the metabolic scale in adult organisms. In their formulations, both approaches treat the scaling exponent $b \approx 3/4$ as a fixed, asymptotic value, characteristic of a ``fully developed'' organism. This assumption, however, shows limitations when confronted with empirical data for organisms undergoing growth. During ontogenetic development, from juvenile stages to maturity, systematic deviations from the three-quarter law are observed, since the effective metabolic exponent is not constant and varies in a predictable way over time. Although the conceptual framework of WBE has been extended to describe ontogenetic growth, it still generally assumes a constant metabolic exponent. Thus, both Kleiber's analysis and the WBE model, by focusing primarily on the relationship between adult individuals of different sizes (an interspecific approach), do not explicitly incorporate the possibility of a dynamic exponent along the life cycle of a single individual. In this respect, these classical models do not fully capture the dynamics of intraspecific scaling during development, leaving a gap precisely in the stages where energy allocation to the construction of new structures (biosynthesis) directly competes with the maintenance of basal functions.

In this context, there arises the need for models that describe the metabolic exponent $b$ not as a universal constant, but as a variable function $b(n)$ of the organism's developmental stage $n$. This limitation of classical models is supported by a growing body of literature documenting variation in the metabolic scaling exponent under different conditions. Glazier shows that the supposed universality of the three-quarter law does not hold either among or within species, with exponents that vary consistently as a function of ontogenetic, ecological, and physiological factors~\cite{Glazier2005,Glazier2010}. This suggests that a more comprehensive theory of metabolic scaling must incorporate the dynamic nature of the exponent $b$ during ontogenetic growth. In light of this evidence, it becomes natural to seek a theoretical framework in which the metabolic exponent $b$ can, in fact, vary throughout development.

Based on the discussion above, we proposed a discrete model in which metabolic activity emerges from an intrinsic logic of developmental programming~\cite{Cambui2025}. In this model, the regulation of metabolism throughout growth is not described as a continuous and homogeneous process, but as a sequence of successive and discrete stages in which the organism reorganizes its energetic priorities, resulting in a distinct effective value for the exponent $b(n)$. Ontogenetic progression is thus understood as a succession of metabolically programmed phases that establishes a formal framework capable of explicitly accommodating the variation of the exponent $b$ as a function of the developmental stage $n$.

Our proposal~\cite{Cambui2025} for an ontogenetic model was developed upon the foundation of Fibonacci dynamics, establishing a mathematical framework that aligns with the observed patterns of biological growth.
In that formulation, the central focus was the analytical derivation of the stage-dependent exponent $b(n)$, its numerical estimation for different mammals, and a systematic comparison with a metabolic database, aiming to test the predictive robustness of the model. In the present article, we adopt a deliberately distinct and complementary perspective. Here, we directly employ the $b(n)$ exponents obtained in that earlier analysis to calculate the corresponding basal metabolic rates via the formula $B(n) = 70\,M^{b(n)}$. Our main objective is not to present new fittings or statistical analyses, but to construct, in a didactic and step-by-step manner, the complete derivation that leads from the recursive growth hypothesis to this functional expression. We thus seek to explicate the internal logic of the model, detail the conceptual interpretation of the ontogenetic dynamics it predicts, and consolidate its theoretical framework, offering a self-contained and pedagogical presentation that was beyond the scope of the previous work.

\begin{center}
	\fbox{\parbox{0.9\textwidth}{
			\textbf{The model in a nutshell:}
			\begin{itemize}
				\item In earlier work, we introduced a discrete ontogenetic model in which
				organisms do not differ only in size, but also in developmental stage
				(e.g., infant, juvenile, adult), labeled by an integer $n$.
				\item In that framework, the metabolic scaling exponent $b$ is not fixed;
				instead, it becomes a stage-dependent quantity $b(n)$.
				\item The stage dependence $b(n)$ arises from a Fibonacci-inspired,
				stepwise growth scheme, in which body mass increases recursively from
				one stage to the next.
				\item In the present paper, we take the exponents $b(n)$ obtained in that
				previous analysis and use them to define a stage-dependent basal
				metabolic rate of the form
				\[
				B(n) = 70\,M^{b(n)},
				\]
				interpreting both $M$ and $b(n)$ in relation to the organism's stage of
				growth.
				\item Within this anchored formulation, Kleiber's three-quarter law
				appears as a special case for stages (or species) in which
				$b(n) \approx 3/4$, while deviations from this value along development
				are made explicit by the function $b(n)$.
			\end{itemize}
	}}
\end{center}

\section{From recursive growth to metabolic scaling}

Before proceeding, it is important to clarify how the present section relates to our earlier work. A formal derivation of the Fibonacci-based ontogenetic model, including the construction of the recursive growth scheme and the analytical expression for the exponent $b(n)$, was already presented in \cite{Cambui2025}, with emphasis on parameter estimation and quantitative comparison with empirical data. Here, we revisit the same mathematical backbone with a different purpose: to reorganize the derivation in a more didactic way and to make explicit how the recursive growth hypothesis leads naturally to a stage-dependent metabolic law of the form $B(n) = B_0 M^{b(n)}$. 

Based on the recursive property of the Fibonacci sequence, expressed by
$F_n = F_{n-1} + F_{n-2}$, we propose a biological analogy. If we imagine that the growth of an organism is not continuous, but occurs in discrete, successive stages (such as juvenile and adult phases), in which the mass acquired at each new stage depends on the gains from the two previous stages, we can transpose this mathematical logic to body mass. This leads to a similar recursive relation for the mass at each stage $n$: $M_n = M_{n-1} + M_{n-2}$. Solving this recurrence, as is done for the Fibonacci sequence, we obtain that the total mass after $n$ stages grows approximately according to a geometric progression, following
\begin{equation}\label{massa}
  M(n) \sim M_0 \,\phi^n,
\end{equation}
where $\phi \approx 1.618$ is the golden ratio.

At first sight, the expression $M(n) \sim M_0 \,\phi^n$ may seem like a bold or even controversial simplification, since it models the growth of the mass of a complex organism using a formula based on the golden ratio, a pattern more frequently associated with static forms than with dynamic processes. However, its value lies less in being a literal and precise description of every gram of tissue, and more in functioning as a minimal conceptual model. It captures the central idea that growth occurs in discrete stages $n$, in which each new step amplifies the existing mass by an approximately constant factor related to $\phi$. Solving Eq.~\eqref{massa} for $n$, we arrive at
\begin{equation}\label{estagio}
  n = \log_\phi\!\left(\frac{M}{M_0}\right),
\end{equation}
which opens the way to relate this stage $n$ to metabolic rate.

The growth stage $n$ allows a fundamental reformulation of the relationship between metabolism and mass. Instead of treating metabolic rate as a static function of mass, we can express it as a dynamic function of the developmental stage:
\begin{equation}\label{Bn}
  B(n) = B_0 \, M^{b(n)},
\end{equation}
in which the scaling exponent $b(n)$ is no longer a constant and now depends explicitly on the ontogenetic stage. This formulation establishes a crucial distinction with respect to the classical law $B = B_0 M^b$. In the classical view, the exponent $b$ is a fixed value, such as Kleiber's $3/4$, which describes an average relationship among adult individuals of different sizes, and mass $M$ plays the role of the only independent variable. In contrast, in the expression $B(n) = B_0 M^{b(n)}$, the exponent $b(n)$ becomes a variable that depends on the developmental stage. This means that, for a single individual, the rule that relates metabolism and mass changes systematically throughout its growth, from the initial stages of rapid mass gain to maturity. While the classical law answers the question "how does an adult elephant differ from an adult rat?", the new formulation seeks to answer "how do the efficiency and metabolic priorities of an organism change as it grows?". In this way, the introduction of the stage $n$ and of the variable exponent $b(n)$ represents a shift from a comparative, static model to a dynamic model that aims to capture the ontogenetic trajectory of metabolism over the life cycle of an individual.

In our model, we propose a mathematical idealization in which the total body mass of an organism at developmental stage $n$, denoted by $M(n)$, grows in proportion to the corresponding term of the Fibonacci sequence, so that $M(n) \sim F_n$ (I). This choice reflects the idea that biological growth often proceeds in discrete steps, in which each new stage incorporates and adds to the structures established in previous stages.

This also captures an important physiological principle: established tissues, even when they are not at maximal functional activity, still require energy to maintain cellular integrity, membrane potential, and routine protein synthesis. This continuous expenditure, essential for maintaining homeostasis and physiological readiness, constitutes the basal maintenance cost. Thus, we have $B(n) \sim F_{n-1}$ (II).

Starting from the general expression for metabolic scaling, $B(n) \sim M(n)^{b(n)}$ (Eq.~\ref{Bn}), we can isolate the stage-dependent scaling exponent, obtaining
\begin{equation}\label{bn1}
  b(n) = \frac{\log B(n)}{\log M(n)}.
\end{equation}
Substituting relations (I) and (II) into Eq.~\eqref{bn1}, we arrive at an expression that links the exponent directly to the ratio between consecutive Fibonacci terms:
\begin{equation}\label{bn-seq}
  b(n) = \frac{\log F_{n-1}}{\log F_n}.
\end{equation}

The Fibonacci sequence has a well-known asymptotic form, given by
\begin{equation}\label{prop1}
  F_n \sim \frac{\phi^n}{\sqrt{5}}, \quad \text{where } \phi = \frac{1+\sqrt{5}}{2},
\end{equation}
and therefore the ratio between consecutive terms satisfies
\begin{equation}\label{prop2}
  \frac{F_{n-1}}{F_n} \sim \frac{\phi^{\,n-1}}{\phi^{\,n}}.
\end{equation}
Inserting the approximations~\eqref{prop1} and~\eqref{prop2} into Eq.~\eqref{bn-seq} and applying logarithmic properties, we obtain the refined form of the exponent:
\begin{equation}\label{bn-ref}
  b(n) = \frac{(n-1)\log \phi - \log\sqrt{5}}{n\log \phi - \log\sqrt{5}}.
\end{equation}

For organisms at advanced developmental stages, where $n$ takes large values, the constant term $\log\sqrt{5}$ becomes negligible compared to $n\log\phi$. In this limit, the expression simplifies significantly to
\begin{equation}\label{bn-simp}
  b(n) \approx \frac{n-1}{n}.
\end{equation}
We refer to Eq.~\eqref{bn-ref} as the refined version of the model, which incorporates the finite-size correction term, and to Eq.~\eqref{bn-simp} as the simplified version, which describes the asymptotic behaviour of the metabolic scaling exponent as a function of the growth stage $n$.

\section{Dynamics of the scaling exponent as a function of stage $n$}

The analytical expressions for the scaling exponent, Eqs.~\eqref{bn-ref} and~\eqref{bn-simp}, transparently determine how $b(n)$ behaves throughout ontogeny. In our previous work~\cite{Cambui2025}, we evaluated these formulas numerically and presented continuous plots of $b(n)$ versus $n$, allowing a detailed comparison with empirical data and with classical baselines such as the WBE exponent $b = 3/4$. For the purposes of this article, it is sufficient to summarize the qualitative features of these trajectories that are most relevant for interpreting the behavior of the metabolic rate $B(n)$.

First, the refined expression~\eqref{bn-ref}, which retains the finite-size correction term $\log\sqrt{5}$, is highly sensitive during the earliest developmental stages. For small values of $n$, the exponent $b(n)$ can deviate significantly from its later trend, even displaying non-monotonic behavior. This is a direct consequence of the discrete, stage-based foundation of the model: at the onset of the Fibonacci-like growth, the constant term $\log\sqrt{5}$ is comparable in magnitude to $n\log\phi$, making the ratio in Eq.~\eqref{bn-ref} strongly influenced by the initial terms of the sequence. Biologically, this mathematical sensitivity suggests that the model should be interpreted with caution when applied to very early ontogenetic phases (e.g., embryonic or larval stages), where the mass--metabolism relationship is known to be governed by highly dynamic and specific developmental processes that a simple scaling law may not fully capture.

Second, as $n$ increases, the situation stabilizes rapidly. Once the organism progresses beyond the most sensitive initial stages, the term $n\log\phi$ dominates both the numerator and denominator of Eq.~\eqref{bn-ref}. Consequently, the refined exponent $b(n)$ converges quickly toward the simplified form $b_{\text{simp}}(n) = (n-1)/n$ given by Eq.~\eqref{bn-simp}. In practice, from moderate values of $n$ onwards, the two versions yield very similar exponents. This indicates that the detailed logarithmic correction primarily affects the early range of development, while the overall long-term trend is governed by the underlying recursive structure of the model.

Finally, both formulations share a key qualitative feature: the exponent $b(n)$ increases monotonically with $n$ and asymptotically approaches $1$ as $n \to \infty$. This implies that, in the theoretical limit of very advanced developmental stages, metabolism tends to scale almost linearly with body mass, that is, $B \propto M^1$. For the intermediate stages relevant to most empirical data, the values of $b(n)$ lie within the well-documented sublinear interval for mammals (approximately between $0.7$ and $0.9$), as discussed in detail in \cite{Cambui2025}. This pattern naturally suggests an ontogenetic transition in which young, rapidly growing organisms operate under stronger efficiency constraints, characterized by a more pronounced sublinear scaling (lower $b(n)$), while mature organisms progressively approach a regime in which the metabolic cost becomes more directly proportional to their total mass (higher $b(n)$). Thus, the stage dependence of $b(n)$ encapsulates, at the level of the scaling exponent itself, how an organism's energetic priorities shift throughout its life cycle.

\section{Kleiber's law as a metabolic anchoring point}

Having revisited the functional form of the exponent $b(n)$ along development, we can now place Kleiber's law within this new ontogenetic framework.

From the standpoint of the proposed model, the metabolic rate thus acquires a dynamic expression that integrates both the universality observed by Kleiber and the variable nature of ontogenetic growth. Starting from the general form
\begin{equation}\label{eq:generalBn}
B(n) = B_{0} \, M^{b(n)} ,
\end{equation}
we can reinterpret the meaning of Kleiber's intercept $B_{0}$. In his classical law, $B \approx 70 \cdot M^{3/4}$, the constant $70$ (in kcal\,day$^{-1}$) represents the expected basal metabolic cost for a hypothetical adult mammal of $1$ kg. In our model, this reference value is not discarded; instead, it acts as a metabolic anchoring point that sets the energetic scale, while the developmental dynamics are captured by the variation of the exponent $b(n)$.

In this way, the expression for the metabolic rate at stage $n$ can be written as
\begin{equation}\label{eq:BnKleiberForm}
B(n) \approx 70 \cdot M^{b(n)} ,
\end{equation}
where $M$ is the total body mass of the organism at that specific stage, and $b(n)$ is given by Eqs.~\eqref{bn-ref} and~\eqref{bn-simp}, depending on whether one wishes to consider the refined correction or the asymptotic limit. This formulation reveals the following dependence: mass $M$ sets the overall scale of the organism, but the scaling rule $b(n)$ is modulated by the developmental stage $n$, which in turn is linked to mass through the approximate relation
\begin{equation}\label{eq:nOfM}
n(M) \approx \log_\phi\!\left(\frac{M}{M_0}\right) .
\end{equation}

Consequently, the model not only recovers Kleiber's law as an \emph{approximate} case, that is, for certain stages in which $b(n) \approx 3/4$, but also provides a framework to systematically describe deviations from this value during growth. In early ontogenetic stages, the model predicts lower values of $b(n)$, which implies a metabolic rate below that predicted by the three-quarter law for a given mass $M$, reflecting a more strongly sublinear scaling regime. As $n$ increases, the exponent $b(n)$ grows and, in the theoretical limit of very advanced stages, tends to $1$, bringing metabolism closer to an almost linear relationship with mass. Thus, Kleiber's intercept remains a fundamental physiological landmark on the energetic scale, while the function $b(n)$ introduces the temporal and developmental dimension that was missing, linking the comparative perspective among species with the dynamic trajectory of each individual over its life cycle.

\section{Ontogenetic results for the metabolic rate $B(n)$ \textit{in mammals}}

In our previous work~\cite{Cambui2025}, we demonstrated the predictive capacity of the Fibonacci-based model by computing the exponents $b(n)$ for nine specific mammalian species, selected based on the availability of reliable data regarding birth and adult mass, and by comparing these exponents with empirical metabolic information from the literature. As stated in Introduction \ref{int}, our objective here is merely to explore the theoretical implications of the model in a more discursive and didactic manner, detailing step by step the derivation of the equations that culminate in the central expression $B(n) \approx 70\,M^{b(n)}$.

We concentrate on analyzing the behavior of this function and the biological interpretation of its dynamics, in particular how the variation of the exponent $b(n)$ modulates the relationship between mass and metabolism throughout development. To this end, and building directly upon the stage-dependent exponents $b_{\text{simp}}(n)$ and $b_{\text{ref}}(n)$ calculated for those nine mammalian species in our previous work~\cite{Cambui2025}, we now compute the corresponding basal metabolic rates via $B(n) = 70\,M^{b(n)}$. The results are summarized in Table~\ref{tab}, which also includes, for comparison, the reference metabolic rate obtained from Kleiber's classical law, $B_{\text{K}} = 70\,M^{3/4}$. When appropriate, we make qualitative comparisons with general trends reported in the literature, illustrating how the theoretical trajectory of $B(n)$ connects with known physiological patterns, without the need for a new exhaustive statistical analysis. In this way, the text aims to consolidate the theoretical foundations of the model and expand its conceptual discussion, establishing a clear bridge between the proposed mathematical structure and the principles of developmental physiology.

\begin{table}[h]
	\centering
	\small
	\caption{Body mass intervals $M$, reference metabolic rates from Kleiber's law ($B_{\text{K}}$), and basal metabolic rate intervals predicted by the ontogenetic model ($B_{\text{simp}}(n)$ and $B_{\text{ref}}(n)$) for nine mammalian species. All metabolic rates are in kcal\,day$^{-1}$. The underlying stage-dependent exponents $b_{\text{simp}}(n)$ and $b_{\text{ref}}(n)$ were obtained in~\cite{Cambui2025}.}
	\label{tab}
	
	\begingroup
	\setlength{\tabcolsep}{7pt} 
	\begin{tabular}{@{}lcccccccc@{}}
		\toprule
		\textbf{Species}& Rabbit & Rat & Mouse & Cat &
		\shortstack{Dog\\(medium)} & Horse & Cow \\
		\midrule
		$M$ (kg) &
		$3.5$--$5.5$ &
		$0.3$--$0.5$ &
		$0.025$--$0.04$ &
		$3.5$--$5.0$ &
		$25$--$35$ &
		$450$--$600$ &
		$600$--$800$  \\
		\bottomrule
	\end{tabular}
	\endgroup
	
		\vspace{0.5em}
	
	\begingroup
	\setlength{\tabcolsep}{7pt} 
	\begin{tabular}{@{}lccc@{}}
		\toprule
		\textbf{Species}/$M$ (kg)&
		 \hspace{1.5cm}Elephant/$4000$--$6300$&
		\hspace{1.05cm}\shortstack{Blue Whale}/$100000$--$150000$ \\
		\bottomrule
	\end{tabular}
	\endgroup

	\vspace{0.5em}
	
	
	\begin{tabular}{@{}l@{\hspace{1.75em}}c@{\hspace{1.75em}}c@{\hspace{1.75em}}c@{}}
		\toprule
		\textbf{Species} & \textbf{$B_{\text{K}}$} & \textbf{$B_{\text{simp}}(n)$} & \textbf{$B_{\text{ref}}(n)$} \\
		\midrule
		Rabbit       & $1.79\times 10^{2}$--$2.51\times 10^{2}$ & $2.09\times 10^{2}$--$3.24\times 10^{2}$ & $2.00\times 10^{2}$--$3.13\times 10^{2}$ \\[4pt]
		Rat          & $2.84\times 10^{1}$--$4.16\times 10^{1}$ & $2.45\times 10^{1}$--$3.80\times 10^{1}$ & $2.56\times 10^{1}$--$3.87\times 10^{1}$ \\[4pt]
		Mouse        & $4.40\times 10^{0}$--$6.26\times 10^{0}$ & $3.29\times 10^{0}$--$4.52\times 10^{0}$ & $4.24\times 10^{0}$--$5.31\times 10^{0}$ \\[4pt]
		Cat          & $1.79\times 10^{2}$--$2.34\times 10^{2}$ & $2.06\times 10^{2}$--$2.83\times 10^{2}$ & $1.95\times 10^{2}$--$2.67\times 10^{2}$ \\[4pt]
		Dog (medium) & $7.83\times 10^{2}$--$1.01\times 10^{3}$ & $1.18\times 10^{3}$--$1.66\times 10^{3}$ & $1.06\times 10^{3}$--$1.53\times 10^{3}$ \\[4pt]
		Horse        & $6.84\times 10^{3}$--$8.49\times 10^{3}$ & $7.31\times 10^{3}$--$1.10\times 10^{4}$ & $2.77\times 10^{3}$--$5.39\times 10^{3}$ \\[4pt]
		Cow          & $8.49\times 10^{3}$--$1.05\times 10^{4}$ & $1.21\times 10^{4}$--$1.91\times 10^{4}$ & $6.74\times 10^{3}$--$1.29\times 10^{4}$ \\[4pt]
		Elephant     & $3.52\times 10^{4}$--$4.95\times 10^{4}$ & $8.99\times 10^{4}$--$1.43\times 10^{5}$ & $6.40\times 10^{4}$--$1.05\times 10^{5}$ \\[4pt]
		Blue Whale   & $3.94\times 10^{5}$--$5.34\times 10^{5}$ & $1.45\times 10^{6}$--$2.42\times 10^{6}$ & $9.02\times 10^{5}$--$1.66\times 10^{6}$ \\
		\bottomrule
	\end{tabular}
	
\end{table}

Table~\ref{tab} summarizes, in a compact form, the main quantities of the model for the nine mammalian species studied. The upper part of the table lists the species and the corresponding body mass intervals $M$ used in the calculations, while the lower part shows, for each of these species (first column), the reference metabolic rate $B_{\text{K}}$ (second column), computed from the classical Kleiber's law and used as a comparative baseline, followed by the basal metabolic rate intervals predicted by the ontogenetic model, $B_{\text{simp}}(n)$ and $B_{\text{ref}}(n)$ (third and fourth columns, respectively), calculated from the relation $B(n) = 70\,M^{b(n)}$ using the stage-dependent exponents determined in~\cite{Cambui2025}. The table allows for a direct visual comparison of how the predictions of the stage-dependent scaling model relate to the classical expectation across a broad range of body sizes.

As discussed in \cite{Cambui2025}, for each species the refined exponent $b_{\text{ref}}(n)$ is systematically smaller than the simplified version $b_{\text{simp}}(n)$, reflecting the influence of the corrective term present in the full expression of the exponent. This difference is small for small and medium-sized species (for example, rat, rabbit, cat, and dog), but becomes progressively more relevant for large species (horse, cow, elephant, and blue whale).

This behavior is directly reflected in the metabolic rate intervals shown in Table~\ref{tab}. For small and intermediate mammals, the ranges of $B_{\text{simp}}(n)$ and $B_{\text{ref}}(n)$ are close to each other and also comparable, in order of magnitude, to the reference rate $B_{\text{K}}$. This indicates that, in these cases, the asymptotic approximation $b_{\text{simp}}(n) = (n-1)/n$ already captures the dependence between mass and metabolism satisfactorily, and that the ontogenetic model reproduces, in a qualitative sense, the predictions of Kleiber's law. In contrast, for large mammals, the metabolic rates based on $b_{\text{ref}}(n)$ are noticeably smaller than those obtained with $b_{\text{simp}}(n)$ and lie closer to $B_{\text{K}}$, revealing an attenuation of the sensitivity of $B(n)$ to mass when the logarithmic correction is taken into account. In biological terms, this means that the refined version avoids overestimating the daily energetic cost of very heavy organisms, yielding predictions more consistent with the idea that metabolic scaling becomes less steep at high body masses.

\begin{figure}[h]
\centering
\includegraphics[width=0.9\textwidth]{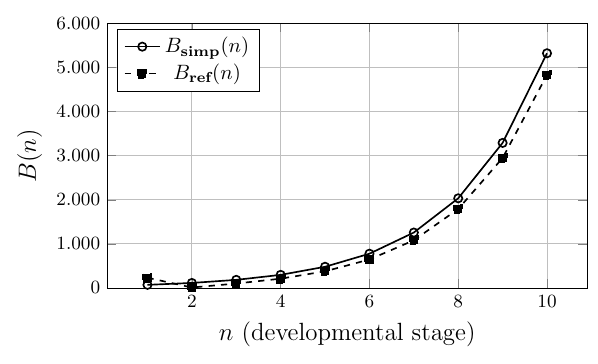}
\caption{Metabolic rate $B(n)$ computed from $B(n)=70\,M(n)^{b(n)}$ for $n=1,\dots,10$, assuming $M(n)\propto\phi^n$. Open circles and solid line denote $B_{\text{simp}}(n)$; filled squares and dashed line denote $B_{\text{ref}}(n)$}
\label{fig1}
\end{figure}

To illustrate the ontogenetic trajectory predicted by the model, Fig.~\ref{fig1} illustrates the behavior of $B(n)$ for $n = 1,\dots,10$ under the idealized assumption $M(n)\propto\phi^n$. The two curves, corresponding to $B_{\text{simp}}(n)$ and $B_{\text{ref}}(n)$, both increase monotonically with the developmental stage, reflecting the higher energetic demands of larger, more developed organisms. The refined trajectory remains slightly below the simplified one, especially at early stages, and gradually converges toward it as $n$ increases, in agreement with the analytical behavior discussed in the previous sections.

In summary, Table~\ref{tab} and Fig.~\ref{fig1} together demonstrate that the model $B(n) = 70\,M^{b(n)}$ generates plausible metabolic rate intervals across several orders of magnitude of body mass and remains consistently aligned with Kleiber's law, which serves here as a classical baseline. At the same time, these results highlight the conceptual role of the two versions of the exponent, given that the simplified form offers a compact and intuitive description of the asymptotic behavior, whereas the refined form introduces an essential correction to more realistically represent the metabolism of large species, bringing its predictions closer to the rates expected from classical scaling.

\section{Conclusion}\label{sec:conclusion}

This work started from a simple yet fundamental observation: organisms are not static, they grow and develop. Classical laws of metabolic scaling, such as Kleiber's law and the WBE theory, were essential in describing \emph{how much} the metabolism of an adult elephant differs from that of an adult rat, but they showed limitations in explaining \emph{how} the metabolism of a single individual changes over the course of its own life, from the juvenile phase to maturity.

To address this gap, we propose a change in perspective. Instead of seeking a single fixed exponent that describes all organisms, we developed a model in which the scaling exponent "grows" together with the organism. Inspired by stage-like growth patterns, so common in nature and mathematically described by the Fibonacci sequence, we structured development as a succession of discrete stages. In this view, the relationship between mass and metabolism is no longer governed by an immutable rule, but is described by a dynamic power law $B(n) = 70\,M^{b(n)}$, in which the exponent $b(n)$ varies systematically with the ontogenetic stage $n$.

The main outcome of this framework is the description of a \emph{metabolic trajectory} along development. The model predicts that, in early stages, energy expenditure scales with mass more gently (lower exponent), which is consistent with the idea that, in these phases, a large fraction of the energy budget is associated with the construction and reorganization of tissues. As the organism matures, the relationship between $B$ and $M$ becomes progressively more direct, with $b(n)$ approaching higher values and, in the limit, a nearly linear regime. In this way, the dynamics of $b(n)$ allow ontogenetic development to be incorporated into scaling theory, using Kleiber's classical value as a metabolic anchoring reference.

This work does not discard the achievements of previous theories, but places them in a broader and more dynamic context. By translating the recursive logic of Fibonacci into the language of physiology, we offer an alternative and complementary narrative for the phenomenon of scaling, intrinsically linked to time and to the process of becoming a complete organism. 

We recognize that the basis of our model, the postulate that mass accumulates according to a geometric progression $M(n) \sim M_0 \cdot \phi^n$ (derived from the Fibonacci recursion), is a fundamental premise, and may seem controversial to some. Within the objectives of this work, we did not identify an equally simple alternative and, in a certain sense, we did not have a conceptual choice that simultaneously preserved stepwise recursion, the connection with the Fibonacci sequence, and the possibility of obtaining an analytical expression for \(b(n)\) and, consequently, for \(B(n)\).  We therefore view this postulate not as a literal description of biological growth, but as a deliberate heuristic that isolates the principle of stage-organized recursion, making visible dynamic relationships that more continuous models may obscure.

We believe that this theoretical framework, centered on principles of gradual organization, paves the way for future investigations into how life history, energy priorities at each stage, and the hierarchical architecture of living beings themselves intertwine to define the rhythm of life.

\end{document}